\documentclass[twocolumn,superscriptaddress,nobibnotes,nofootinbib]{revtex4-2}
\usepackage[utf8]{inputenc}
\usepackage{amsmath}
\usepackage[english]{babel}
\usepackage[dvipsnames]{xcolor}
\usepackage{amssymb}
\usepackage{textcomp}
\usepackage{graphicx}
\usepackage{placeins}
\usepackage{textgreek}
\usepackage{upgreek}
\usepackage{latexsym}
\usepackage{braket}
\usepackage[nolist]{acronym}
\usepackage{siunitx}
\usepackage{xpatch}
\usepackage{dsfont}
\usepackage{hyperref}
\usepackage[dvipsnames]{xcolor}
\usepackage{soul}

\setcitestyle{super} %citations as super script

\hypersetup{linktoc=all, hyperindex=true,colorlinks, linkcolor={red!50!black},citecolor={blue!50!black},urlcolor={blue!80!black}}

\makeatletter

\def\l@subsubsection#1#2{}
\patchcmd{\@ssect@ltx}
    {\addcontentsline{toc}{#1}{\protect\numberline{}#8}}
    {}
    {}
    {}
\makeatother

\begin{document}
\newacro{lqg}[LQG]{linear quadratic gaussian regulator}
\newacro{lqr}[LQR]{linear quadratic regulator}

\newcommand{\bop}{\ensuremath{b}}
\newcommand{\bdag}{\ensuremath{b^{\dagger}}}
\newcommand{\aop}{\ensuremath{a}}
\newcommand{\adag}{\ensuremath{a^{\dagger}}}
\newcommand{\cop}{\ensuremath{c}}
\newcommand{\cdag}{\ensuremath{c^{\dagger}}}
\newcommand{\dd}[1]{\ensuremath{\mathrm{d}#1\,}}
\newcommand{\ii}{\ensuremath{\mathrm{i}}}
\newcommand{\ee}{\ensuremath{\mathrm{e}}}
\newcommand{\ddd}[1]{\ensuremath{\mathrm{d}^3#1\,}}
\newcommand{\mean}[1]{\ensuremath{\langle  #1 \rangle}}
\newcommand{\vinvis}{\rule{0pt}{4ex}}
\newcommand{\negvinvis}{\rule[-4ex]{0pt}{4ex}}
\newcommand{\hoverbrace}[2]{\ensuremath{\overbrace{\vinvis #1}^{#2}}}
\newcommand{\hunderbrace}[2]{\ensuremath{\underbrace{\negvinvis #1}_{#2}}}

%
%title
\title{Squeezed light from a levitated nanoparticle at room temperature}

\newcommand{\vcq}{University of Vienna, Faculty of Physics, Vienna Center for Quantum Science and Technology (VCQ), 1090 Vienna, Austria}

\newcommand{\iqoqi}{Institute for Quantum Optics and Quantum Information (IQOQI), Austrian Academy of Sciences, 1090 Vienna, Austria.}

%
%authors
\author{Lorenzo Magrini}
\email{lorenzo.magrini@univie.ac.at}
\affiliation{\vcq}
\author{Victor A. Camarena-Ch\'{a}vez}
\affiliation{\vcq}
\author{Constanze Bach}
\affiliation{\vcq}
\author{Aisling Johnson}
\affiliation{\vcq}
\author{Markus Aspelmeyer}
\email{markus.aspelmeyer@univie.ac.at}
\affiliation{\vcq}
\affiliation{\iqoqi}

\begin{abstract}

Quantum measurements of mechanical systems can produce optical squeezing via ponderomotive forces. Its observation requires high environmental isolation and efficient detection, typically achieved by using optical cavities and cryogenic cooling. Here we realize these conditions by measuring the position of an optically levitated nanoparticle at room temperature and without the overhead of an optical cavity. We use a fast heterodyne detection to reconstruct simultaneously orthogonal optical quadratures, and observe a noise reduction of $9\%\pm0.5\%$ below shot noise. Our experiment offers a novel, cavity-less platform for squeezed-light enhanced sensing. At the same time it delineates a clear and simple strategy towards observation of stationary optomechanical entanglement.

\end{abstract}

\flushbottom
\maketitle

\thispagestyle{empty}

\section*{Introduction}

A series of recent experiments has demonstrated position measurements of optically levitated nanoparticles close to the Heisenberg limit~\cite{magrini2021, Tebbenjohanns2021}. In this regime the measurement process is dominated by the quantum backaction, i.e. radiation-pressure shot noise, of the probe laser. In combination with high detection efficiency this allows for real-time motional quantum control of the particle. At the same time, however, the measurement process also affects the optical probe beam: radiation-pressure induced position fluctuations generate correlations between the amplitude- and phase quadrature of the scattered light. This results in (ponderomotive) optical squeezing, in which the noise of a field quadrature can be below the vacuum shot noise~\cite{fabre1994ponderomotive,mancini1994ponderomotive}. Thus far, the observation of this effect has required optical cavities~\cite{Brooks2012,Purdy2017, Sudhir2017rt,Yu2020,Aggarwal2020} to enhance both measurement strength and detection efficiency, often combined with cryogenic cooling~\cite{Marino2010, Verlot2010,SafaviNaeini2013,Purdy2013, Sudhir2017,Ockeloen2018, Mason2019} to reduce thermal noise contributions. In this work we show ponderomotive squeezing in its simplest form: at room temperature and without an optical cavity. Using heterodyne detection allows us to provide a complete tomography of the optical state, which is a crucial step towards measuring optomechanical entanglement in such a system~\cite{gut2020}. Given the relevance of optical squeezing for precision measurements~\cite{Schnabel2017,Yu2020,Taylor2013b, Casacio2021}, our experiment also offers an interesting platform for cavity-less quantum sensing applications.

\section*{Experiment}

\begin{figure}[!hb]
\centering
\includegraphics[width = 8cm]{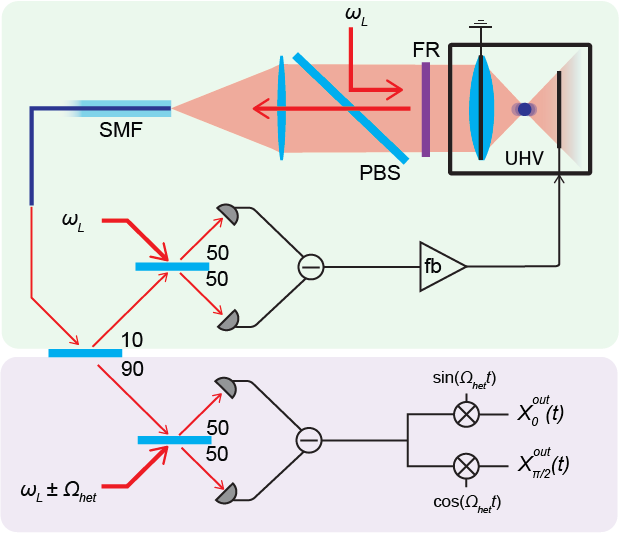}
\caption{\textbf{Experimental setup}. A silica particle is trapped in an optical tweezer (optical frequency: $\omega_L$) in ulta high vacuum (UHV). The back-scattered light is isolated by a faraday rotator (FR) in combination with a polarizing beam splitter (PBS), and focussed on to a single mode fiber (SMF) as part of a fiber based confocal microscope. A beamspitters with splitting ratio of 10:90 distributes the light between homodyne and heterodyne detectors. The homodyne measurement is used for electrostatic feedback (fb). The heterodyne signal is numerically demodulated to obtain orthogonal quadrature components.}
\label{fig:1}
\end{figure}

A silica nanoparticle (radius $r=43\,\mathrm{nm}$) is trapped in an optical tweezer ($\mathrm{NA}=0.95$) in ultra high vacuum ($\sim 10^{-8}\, \mathrm{mbar}$). At a wavelength of $\lambda_0 = 1064\, \mathrm{nm}$ and a power of $P \sim 150 \,\mathrm{mW}$, the trapping frequency along the tweezer axis is $\Omega_z/2\pi = 76.9 \,\mathrm{kHz}$. The transverse frequencies, are $\Omega_x/2\pi = 178 \,\mathrm{kHz}$ (perpendicular to the direction of polarization) and $\Omega_y/2\pi = 229 \,\mathrm{kHz}$ (parallel to the direction of polarization). We use a fiber-based confocal microscope to collect light that is scattered back from the particle. This allows to maximize detection efficiency~\cite{Tebbenjohanns2019_detection} while suppressing scattering of stray light into the detector~\cite{Vamivakas2007, magrini2021}. A small fraction (10\%) of the back-scattered light is sent to a homodyne receiver and is used to cool the axial mode by applying linear feedback  in form of an electrostatic force~\cite{Frimmer2017, Tebbenjohanns2019} (Figure \ref{fig:1}). The transverse modes are stabilized by parametric modulation of the trap stiffness~\cite{Gieseler2012, magrini2021}.
The remaining 90\% of the collected back-scattered light represents the center of this study. We detect this on a heterodyne receiver with a local oscillator shifted by $\Omega_\mathrm{het}/2\pi =6\,\mathrm{MHz}$ by using a sequence of acousto-optic modulators. This allows us to reconstruct simultaneously the amplitude ($X^\mathrm{out}_{0}$) and phase ($X^\mathrm{out}_{\frac{\pi}{2}}$) quadratures of the measured field, from which, by linear combination, any quadrature can be derived:
\begin{equation}\label{eq:quadratures}
X^{\mathrm{out}}_{\theta}  =X^{\mathrm{out}}_{0}\cos(\theta) + X^{\mathrm{out}}_{\frac{\pi}{2}}\sin(\theta),
\end{equation}
A simultaneous measurement of the two orthogonal quadratures could be equivalently obtained by using two homodyne receivers with a fixed $\pi/2$ phase relation (dual-rail homodyning)~\cite{walker1987}. The advantage of a heterodyne measurement is that power monitoring and active phase stabilization of the interferometer are not required~\cite{magrini_thesis2021}.

The squeezing of noise below vacuum, originating from the optomechanical interaction, is evident in the spectrum of the detected quadratures $X^{\mathrm{out}}_{\theta}$, which can be written as~\cite{supplementary}:
\begin{equation}\label{eq:ponderomotive1}
\begin{split}
S_{\theta,\theta}^{\mathrm{det}}(\Omega) = 1 + \frac{4\Gamma_{\mathrm{meas}}}{z_{\mathrm{zpf}}^2} 
&\left[ \vphantom{\frac{1}{1}} S_{zz}(\Omega)\sin^2(\theta) \right. \\
&\left. +\frac{\hbar}{2}\mathrm{Re}\left\lbrace\chi_\mathrm{m}(\Omega)\right\rbrace\sin(2\theta)\right].
\end{split}
\end{equation}
Three contributions to equation \eqref{eq:ponderomotive1} can be clearly distinguished. The first term represents the vacuum noise of the input field, to which the spectrum is normalized. The second contribution depends on the mechanical displacement noise $S_{zz}(\Omega) =\left(S_{FF}^\mathrm{ba}+S_{FF}^\mathrm{th}\right) \lvert\chi_\mathrm{m}(\Omega)\rvert^2$, that is given by the response of the mechanical oscillator to the measurement backaction and thermal force noise contributions ($S_{FF}^\mathrm{ba,th}$) via the mechanical susceptibility $\chi_\mathrm{m}(\Omega)$. The force noise contributions can also be conveniently expressed in terms of theis corresponding decoherence rates $ \Gamma_\mathrm{ba,th}=S_{FF}^\mathrm{ba,th}/(4p^2_\mathrm{zpf})$, with $p_\mathrm{zpf}=\sqrt{\hbar m\Omega_z/2}$ being the momentum zero-point fluctuation. In this context, the prefactor $\Gamma_\mathrm{meas} = \eta_\mathrm{d} \Gamma_\mathrm{ba}$ ($\eta_\mathrm{d}$: detection efficiency) denotes the measurement rate, representing the inverse of the time required to resolve the zero-point motion ($z_\mathrm{zpf}= \sqrt{\hbar/(2m\Omega_z)}$) of the oscillator~\cite{Wilson2015}. Finally, the third term describes the correlation between the phase and amplitude of the scattered optical field. It originates in the optomechanical interaction between the light and the particle, where backaction induced position fluctuations result in correlated phase fluctuations of the backscattered light~\cite{supplementary}. While the first two terms are necessarily positive, the third can contribute with a negative sign, allowing, for an appropriate quadrature angle and frequency range, a noise reduction below the vacuum fluctuations, i.e. ponderomotive squeezing.

\section*{Results}

\begin{figure}[h!]
\centering
\includegraphics[scale=1]{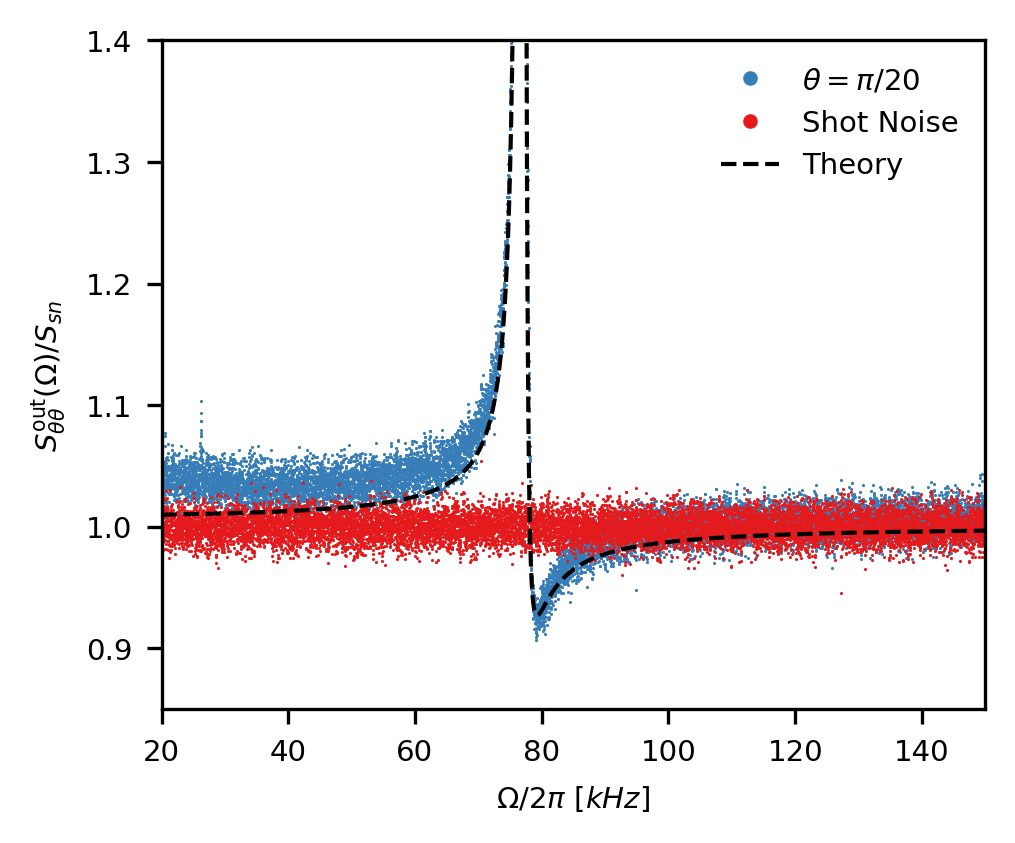}
\caption{\textbf{Ponderomotive squeezing}. Shown is the spectrum of the optical quadrature at an angle $\theta = \pi/20$  (blue) and the optical shot noise of the heterodyne measurement (red), which is independently measured by blocking the signal path. Close to resonance we observe a noise reduction of about $9\%$ below shot noise. The black dashed line shows the theoretical plot expected for our experimental conditions.}
\label{fig:2}
\end{figure}

We record a 500 second heterodyne trace of the light scattered by the particle, and reconstruct the two orthogonal quadratures by using a Hilbert transform signal decomposition. This gives us access, using definition \eqref{eq:quadratures}, to the simultaneous time record of the optical quadratures for any given angle $\theta$. We observe a maximum squeezing of $9\%\pm0.5\%$ (without dark noise subtraction) at a quadrature angle $\theta\sim \pi/20$ (Figure \ref{fig:2}). This is consistent with the optimal value obtained by minimizing equation~\ref{eq:ponderomotive1}. The uncertainty is given by the $95\%$ confidence interval of the expected $\chi^2$ distribution of the averaged  spectrum~\cite{Wieczorek2015}.
\begin{figure*}[ht!]
\includegraphics[scale =1]{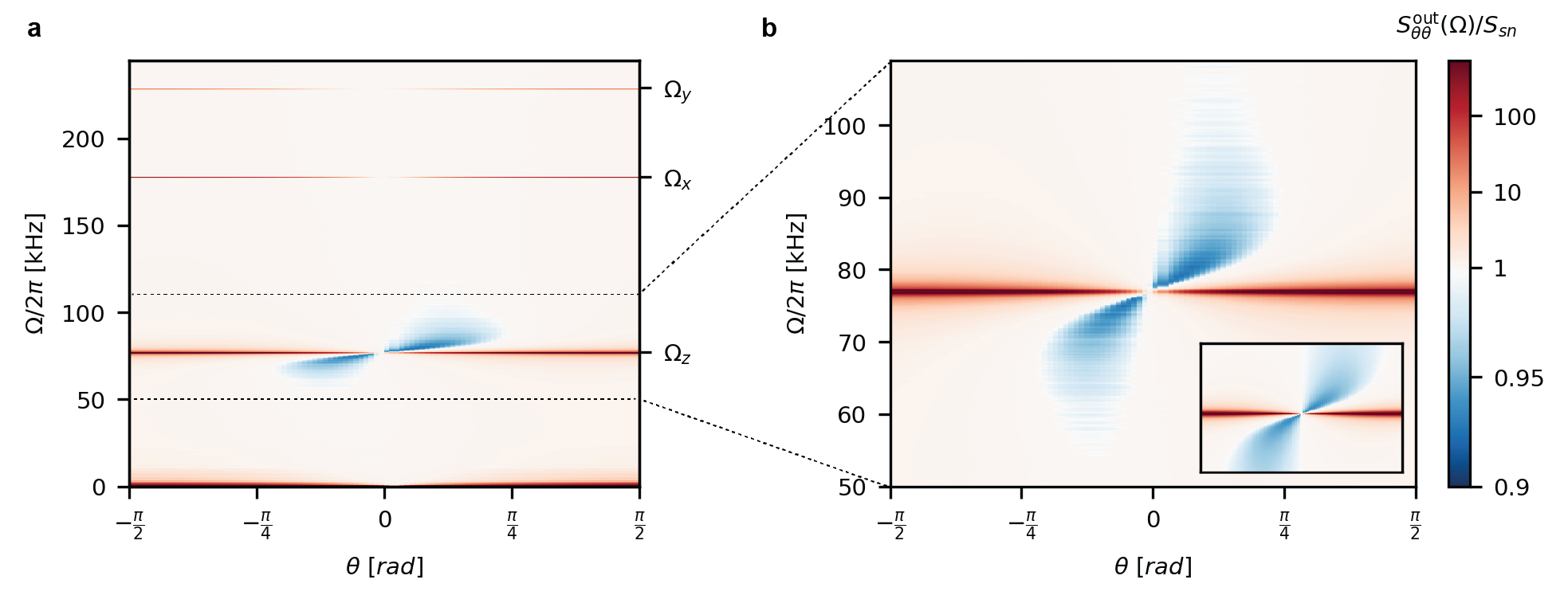}
\caption{\textbf{Ponderomotive spectrogram.} Spectrogram of the measured output quadratures as a function of the analyzer angle, where the warm and cold colours represent noise values above and below shot noise, respectively. \textbf{a}. The signal at $\Omega\sim0$ arises from the low frequency phase noise (minimized at $\theta=0$). For $\Omega =2\pi=76.9\, \mathrm{kHz}$, the negative contribution of the squeezed light outweighs the other noise contributions in this frequency range, taking the spectrum below the shot noise level, in accordance with equation \ref{eq:ponderomotive1}. No squeezing can be observed for the transverse modes at $\Omega_x$ and $\Omega_y$. \textbf{b} Detail of the measured spectrogram around the main resonance $\Omega_z$. The inset shows the theoretical plot expected for our experimental conditions.}\label{fig:3}
\end{figure*}
Equation \eqref{eq:ponderomotive1} is strongly nonlinear in its parameters (in particular in $\Gamma_\mathrm{ba}$) and hence can make fitting the model \eqref{eq:ponderomotive1} to the measured autocorrelation spectrum $S_{\theta,\theta}(\Omega)$ a non trivial task. For this reason we take advantage of the properties of the cross correlation spectrum $S_{\frac{\pi}{4},\frac{3\pi}{4}}(\Omega)$ (readily available from our heterodyne measurement) which allows to access a calibration-free estimate of the steady state energy of the harmonic oscillator in units of $\hbar\Omega_z$~\cite{Purdy2017}. From the ratio of the real and imaginary part of $S_{\frac{\pi}{4},\frac{3\pi}{4}}(\Omega)$ we estimate a steady state occupation of $n=84.9\pm 0.7$\cite{supplementary}. With a feedback dominated damping of $\gamma/2\pi = 89.7\pm 0.3\, \mathrm{Hz}$, the total decoherence rate of our system is $\Gamma_\mathrm{tot} = \gamma n = 2\pi (7.6 \pm 0.1\, \mathrm{kHz})$\cite{supplementary}. We can use this to fit $S_{\frac{\pi}{2},\frac{\pi}{2}}(\Omega)$ (where the squeezing term vanishes) with only $\Gamma_\mathrm{meas}$ as free parameter, and obtain $\Gamma_\mathrm{meas}/2\pi = 0.6\pm 0.1\, \mathrm{kHz}$. We then derived the expected squeezing spectra $S_{\theta,\theta}(\Omega)$, and observe excellent agreement with our data (Figures \ref{fig:2} and \ref{fig:3}). The slight discrepancy at low frequencies is due to excess phase noise caused by the signal generators that are used to produce the heterodyne local oscillator. At our experimental pressure of $1.5\cdot10^{-8}\,\mathrm{mbar}$ we estimate a thermal decoherence rate of $\Gamma_\mathrm{th}/2\pi = 2.7, \mathrm{kHz}$, resulting in a measurement cooperativity of $C_\mathrm{q}=\Gamma_\mathrm{ba}/\Gamma_\mathrm{th} = 1.8$, and in a detection efficiency per quadrature of $\eta_\mathrm{d} = 0.11$.

\section*{Discussion and outlook}

In conclusion, we have demonstrated ponderomotive squeezing of light in a simple experimental setting without requiring a cavity and at room temperature. The agreement between the measured data and the theoretical description is excellent. In general, the amount of obtainable squeezing depends on both the cooperativity $C_\mathrm{q}$ and the detection efficiency $\eta_\mathrm{d}$, which have not been pushed to the extreme in our proof of concept demonstration. Furthermore, our heterodyne measurement strategy reduces the largest possible squeezing by a factor of 2, since the measurement information is shared between both quadratures. On the other hand, however, heterodyning provides us with full state tomography of the optical state. Looking into the future, it also allows us to reconstruct the full optomechanical covariance matrix and search for quantum correlations between light and mechanics to reveal stationary  entanglement of the mechanical oscillator with its measurement device, the light field~\cite{gut2020}.

\paragraph*{Note added.} We recently became aware of a related independent work by Rossi \textit{et al.}.

\section*{Acknowledgements}
We thank Nikolai Kiesel, Sebastian G. Hofer, Klemens Winkler and Corentin Gut for the helpful discussions.
This project was supported by the European Research Council (ERC 6 CoG QLev4G), by the ERA-NET programme QuantERA under the Grants QuaSeRT and TheBlinQC (via the EC, the Austrian ministries BMDW and BMBWF and research promotion agency FFG), by the European Union’s Horizon 2020 research and innovation programme under Grant No. 863132 (iQLev). L.M. and V.A.C.C. are supported by the Vienna Doctoral School of Physics (VDS-P) and L.M. is supported by the FWF under project W1210 (CoQuS).

\FloatBarrier
\let\oldaddcontentsline\addcontentsline% Store \addcontentsline
\renewcommand{\addcontentsline}[3]{}% Make \addcontentsline a 
\bibliographystyle{bibliography_nature}
\bibliography{Bibliography}
\let\addcontentsline\oldaddcontentsline% Restore \addcontentsline

%%%%%%%SUPPLEMENTARY%%%%%%%%%%%%%%%%%%

\clearpage
\onecolumngrid

\renewcommand{\thetable}{A\arabic{table}}  
\renewcommand{\thefigure}{A\arabic{figure}} 
\renewcommand{\theHfigure}{A\arabic{figure}}
\renewcommand{\theequation}{A\arabic{equation}} 
\setcounter{figure}{0}
\setcounter{equation}{0}
\setcounter{section}{0}
\setcounter{table}{0}

\begin{center}
\large\textbf{\uppercase{Appendix}}
\end{center}

\vspace{3,5cm}
\tableofcontents
\vspace{3,5cm}

\section{Theory}

In the following, we derive the time dependence of the output optical field after its interaction with a mechanical oscillator~\cite{magrini_thesis2021}. This is conveniently done in the Heisenberg picture, with the quantum operators evolve in time, according to the Heisenberg equation:
\begin{equation}\label{eq:ch1:heiseq}
\dot{q}_i(t) = \frac{\ii}{\hbar}\left[H,q_i(t) \right]
\end{equation}
The detailed derivation of the quantum Langevin equations of motion of the optical and mechanical creation and annihilation operators presented here is equivalent to the derivation showed in Magrini \textit{et al.}~\cite{magrini2021} (Supplementary Material). However, for the sake of simplicity spatial mode considerations will be neglected, absorbing the effects of imperfect overlap of the output and detection modes into an overall measurement efficiency. This is justified by the final results of~\cite{magrini2021}, and allows us to focus the attention on the important mathematical steps and remain pedagogical without losing sight of the general picture.

After having derived the time-dependent equations of motion of the mechanical oscillator and the output optical quadratures, we will analyze these in frequency space, where the main approximations are considered and the effects of ponderomotive squeezing are most easily represented.

\subsection{The quantum Langevin equations}

We start by defining the Hamiltonian of the optomechanical system: a mechanical oscillator (defined by the operators $\aop$, $\adag$)  dissipatively coupled to an external thermal environment or bath (defined by the operators $\bop$, $\bdag$) and dispersively coupled to a measurement apparatus or optical field (defined by the operators $\cop$, $\cdag$):
%\marginpar{Notation: the frequency mode label of the creation and annihilation operators is omitted.}
\begin{equation}
\begin{split}
H &= \hoverbrace{\hbar\Omega_q \adag\aop}{\mathrm{system}}
+ \hoverbrace{\hbar \int \dd{\omega}\omega\,\bdag\bop}{\mathrm{bath}}
+ \hoverbrace{\hbar \int \dd{\omega}\omega\,\cdag\cop}{\mathrm{meter}}
\\ 
&+\hunderbrace{\ii\hbar\sqrt{\frac{\gamma}{2\pi}} \int \dd{\omega}\left(\bdag\aop+\bop\adag\right)}{\mathrm{thermal \quad coupling}}
+\hunderbrace{\hbar g \int \dd{\omega}\left(\adag+\aop\right)\left(\cdag+\cop\right)}{\mathrm{measurement\quad interaction}}
\end{split}
\end{equation}
where the optomechanical coupling $g$ and damping $\gamma$ rates are assumed to be real and independent of frequency.
%\marginpar{Math reminder:\\$[\adag\aop,\aop] = \adag\aop\aop - \aop\adag\aop = [\adag,\aop]\aop +\aop\adag\aop-\aop\adag\aop =[\adag,\aop]\aop = -\aop$}
We can now write the the Heisenberg equations of motion \eqref{eq:ch1:heiseq} for $\aop$, $\bop$ and $\cop$:
\begin{subequations}
\begin{align}
\label{eq:ch5:heis_a}
\dot{\aop}(t)&= \frac{\ii}{\hbar}\left[H ,\aop \right] = -\hoverbrace{ \ii\Omega_q \aop(t)}{A1}
-\hoverbrace{ \sqrt{\frac{\gamma}{2\pi}}\int\dd{\omega}\bop(t)}{B1}
+\hoverbrace{\ii g\int\dd{\omega}\left(\cdag(t)+\cop(t)\right)}{C1}
\\
\label{eq:ch5:heis_b}
\dot{\bop}(t)&=\frac{\ii}{\hbar}\left[H ,\bop \right] = -\ii\omega\bop(t)+\sqrt{\frac{\gamma}{2\pi}}\aop
\\
\label{eq:ch5:heis_c}
\dot{\cop}(t)&=\frac{\ii}{\hbar}\left[H ,\cop \right]= -\ii\omega\cop(t)-\ii g\left(\adag+\aop\right)
\end{align}
\end{subequations}

\subsection{The system evolution}

We now formally solve the linear differential equations \eqref{eq:ch5:heis_b} and \eqref{eq:ch5:heis_c} by integrating them up to the interaction time $t$. These solutions can then be used in terms of $B$ and $C$ in equation \eqref{eq:ch5:heis_a}. Here it is convenient to recognize (and define) the input modes for the thermal and optical interaction, which allow us to simplify the notation. 
Equations \eqref{eq:ch5:heis_b} and \eqref{eq:ch5:heis_c} are a couple of  linear, first order, differential equations, their solution is:
%\marginpar{Math reminder:\\$\dot{x}(t)=\lambda x(t)+f(t)$,\\substitution: $y(t) = x\ee^{-\lambda t}$,\\derivative: $\dot{y}(t) = (x(t)-\lambda)\ee^{-\lambda t}~=~f(t)\ee^{-\lambda t}$\\solution: $y(t) = y(0)+\int\limits_0^t \dd{t^\prime}\ee^{-\lambda t^\prime}f(t\prime)$\\substitute back: $x(t) =x(0)\ee^{\lambda t}+\int\limits^t_0 \dd{t}\ee^{\lambda(t-t^\prime)}f(t^\prime)$}
\begin{subequations}
\begin{align}
\label{eq:ch5:heis2_b}
\bop(t) &= \bop(0)\ee^{-\ii\omega t}+\sqrt{\frac{\gamma}{2\pi}}\int\limits_0^t \dd{t^\prime}\ee^{-\ii\omega(t-t^\prime)}\aop(t^\prime)\\
\label{eq:ch5:heis2_c}
\cop(t) &= \cop(0)\ee^{-\ii\omega t}-\ii g\int\limits_0^t \dd{t^\prime}\ee^{-\ii\omega(t-t^\prime)}\left(\aop^\dagger(t^\prime)+\aop(t^\prime)\right)
\end{align}
\end{subequations}
Now we can plug equation \eqref{eq:ch5:heis2_b} into the $B1$ term of equation \eqref{eq:ch5:heis_a}:
%\marginpar{\vspace{3cm}\\ Math reminder:\\$\int\limits_a^b\dd{x}\delta(x-b)f(x) = \frac{1}{2}f(b)$}
\begin{equation}\label{eq:ch5:heis3_b}
\begin{split}
B1 &= \sqrt{\frac{\gamma}{2\pi}}\int \dd{\omega} \bop(t)\\
&=  \sqrt{\frac{\gamma}{2\pi}}\int \dd{\omega} \left(\bop(0)\ee^{-i\omega t}+\sqrt{\frac{\gamma}{2\pi}}\int\limits_0^t \dd{t^\prime}\ee^{-i\omega(t-t^\prime)}\aop(t^\prime) \right)\\
&=  \sqrt{\gamma}\hunderbrace{\sqrt{\frac{1}{2\pi}}\int \dd{\omega} \bop(0)\ee^{-i\omega t}}{-\bop_\mathrm{in}(t)} +\frac{\gamma}{2\pi}\int\limits_0^t \dd{t^\prime}\aop(t^\prime)\hunderbrace{\int\dd{\omega}\ee^{-\ii\omega(t-t^\prime)}}{2\pi\delta(t-t^\prime)}\\
&= -\sqrt{\gamma}\bop_\mathrm{in}(t)+\frac{\gamma}{2}\aop(t)
\end{split}
\end{equation}
where we have defined the input mode of the thermal bath:
%\marginpar{Properties~\cite{gardiner_input_1985}\\$\mean{\bop_\mathrm{in}(t)\bop_\mathrm{in}(0)}=0$\\$\mean{\bdag_\mathrm{in}(t)\bdag_\mathrm{in}(0)}=0$\\$\mean{\bdag_\mathrm{in}(t)\bop_\mathrm{in}(0)}=n_\mathrm{th}\delta(t)$\\$\mean{\bop_\mathrm{in}(t)\bdag_\mathrm{in}(0)}=(n_\mathrm{th}+1)\delta(t)$}
\begin{equation}
\bop_\mathrm{in}(t) = -\sqrt{\frac{1}{2\pi}}\int \dd{\omega} \bop(0)\ee^{-i\omega t}\,.
\end{equation}
We proceed similarly with the optical field: we use \eqref{eq:ch5:heis2_c} into the $C1$ term of equation \eqref{eq:ch5:heis_a}:
\begin{equation}\label{eq:ch5:heis3_c}
\begin{split}
C1 &= \ii g\int \dd{\omega} \left(\cop^\dagger(t) + \cop(t)\right)\\
&=  \ii g\int \dd{\omega} \left[\left(\cop^\dagger(0)\ee^{i\omega t}+\ii g\int\limits_0^t \dd{t^\prime}\ee^{i\omega(t-t^\prime)}\left(\aop^\dagger(t^\prime)+\aop(t^\prime)\right) \right)+\mathrm{H.c.}\right]\\
&= \ii g\biggl(\hunderbrace{\int \dd{\omega} \cop^\dagger(0)\ee^{\ii\omega t}}{\sqrt{2\pi}\cop_\mathrm{in}^\dagger(t)}+\hunderbrace{\int \dd{\omega} \cop(0)\ee^{-\ii\omega t}}{\sqrt{2\pi}\cop_\mathrm{in}(t)}\biggr)\\&\quad\quad\quad\quad\quad\quad-g^2\int\limits_0^t \dd{t^\prime} \left(\aop^\dagger(t^\prime)+\aop(t^\prime)\right) \hunderbrace{\int \dd{\omega} \ee^{\ii\omega (t-t^\prime)} - \ee^{-\ii\omega (t-t^\prime)}}{2\pi\delta(t-t^\prime)-2\pi\delta(t-t^\prime)=0}\\
&=\ii \sqrt{2\pi} g \left(\cop_\mathrm{in}^\dagger(t)+\cop_\mathrm{in}(t)\right)
\end{split}
\end{equation}
where we have defined the input mode of the optical field, up to the
%\marginpar{Properties~\cite{Loudon}:\\$\mean{\cop_\mathrm{in}(t)\cop_\mathrm{in}(0)}=0$\\$\mean{\cdag_\mathrm{in}(t)\cdag_\mathrm{in}(0)}=0$\\$\mean{\cdag_\mathrm{in}(t)\cop_\mathrm{in}(0)}=0$\\$\mean{\cop_\mathrm{in}(t)\cdag_\mathrm{in}(0)}=\delta(t)$}
interaction time $t$ as:
\begin{equation}
\cop_\mathrm{in}(t) = \sqrt{\frac{1}{2\pi}}\int \dd{\omega} \cop(0)\ee^{-i\omega t} \,.
\end{equation}
We can finally write the evolution of the $\aop$ operator inserting the results for $B1$ \eqref{eq:ch5:heis3_b} and $C1$ \eqref{eq:ch5:heis3_c} into equation \eqref{eq:ch5:heis_a}:
\begin{equation}
\dot{\aop}(t) =  -\left(\ii\Omega_q+\frac{\gamma}{2}\right) \aop(t) +\sqrt{\gamma}\bop^\mathrm{in}(t) + \ii \sqrt{\Gamma_\mathrm{ba}}\left(\cop_\mathrm{in}^\dagger(t)+\cop_\mathrm{in}(t)\right)
\end{equation}
where we have defined the backaction induced decoherence rate as $\Gamma_\mathrm{ba} = 2\pi g^2$. The solution of this linear differential equation is conveniently solved in the Fourier domain (note that we now explicitly write the frequency dependence of the mode operators).
%\marginpar{Math reminder:\\ $\mathcal{F}[\aop(t)]=\aop(\omega)=\sqrt{\frac{1}{2\pi}} \int \dd{t}\ee^{-i\omega t}\aop(t)$\\$ $\\$\mathcal{F}^{-1}[\aop(\omega)]=\aop(t)=\sqrt{\frac{1}{2\pi}}\int \dd{t}\ee^{i\omega t}\aop(\omega)$\\$ $\\ $\dot{\aop}(\omega) = -i\omega \aop(\omega)$\\ $(\aop(\omega))^\dagger = \adag(-\omega)$} 
This is simply done by writing the operators in terms of their Fourier components, allowing to apply the time derivative to the exponential term only. This gives:
\begin{subequations}\label{eq:ch5:aomega}
\begin{align}
\aop(\omega) &= \frac{\sqrt{\gamma} \bop_\mathrm{in}(\omega)+ \ii\sqrt{\Gamma_\mathrm{ba}} \left(\cdag_\mathrm{in}(\omega) +  \cop_\mathrm{in}(\omega)\right)}{ \ii \left( \Omega_q - \omega \right) +\frac{\gamma}{2}}\\
\adag(\omega) &= \frac{\sqrt{\gamma} \bdag_\mathrm{in}(\omega)- \ii\sqrt{\Gamma_\mathrm{ba}} \left(\cdag_\mathrm{in}(\omega)+  \cop_\mathrm{in}(\omega)\right)}{- \ii \left( \Omega_q + \omega \right) +\frac{\gamma}{2}}
\end{align}
\end{subequations}
and the correlators of the bath and input field in Fourier space become~\cite{Hauer2015}:
\begin{subequations}\label{eq:ch5:correlators}
\begin{align}
\begin{split}
\mean{\bop_\mathrm{in}(\omega)\bop_\mathrm{in}(\omega^\prime)}=\mean{\bdag_\mathrm{in}(\omega)\bdag_\mathrm{in}(\omega^\prime)}&=0\\
\mean{\bdag_\mathrm{in}(\omega)\bop_\mathrm{in}(\omega^\prime)}&=n_\mathrm{th}\delta(\omega+\omega^\prime)\\
\mean{\bop_\mathrm{in}(\omega)\bdag_\mathrm{in}(\omega^\prime)}&=\left(n_\mathrm{th}+1\right)\delta(\omega+\omega^\prime)
\end{split}\\
\begin{split}
\mean{\cop_\mathrm{in}(\omega)\cop_\mathrm{in}(\omega^\prime)}=\mean{\cdag_\mathrm{in}(\omega)\cdag_\mathrm{in}(\omega^\prime)}=\mean{\cdag_\mathrm{in}(\omega)\cop_\mathrm{in}(\omega^\prime)}&=0\\
\mean{\cop_\mathrm{in}(\omega)\cdag_\mathrm{in}(\omega^\prime)}&=\delta(\omega+\omega^\prime)
\end{split}
\end{align}
\end{subequations}

\subsection{The output field}

We now consider the output mode of the optical field, which is the experimentally accessible quantity. We can define the output optical field in a similar way to what was done for the input optical field in equation \eqref{eq:ch5:heis2_c} as a function of the field operators at the moment of interaction. However, rather than defining an earlier time $0$ and integrating up to the interaction time $t$, we define a late time $T$, and integrate backwards in time, back to the instant of interaction. Equation \eqref{eq:ch5:heis2_c} can equivalently be written as:
\begin{equation}\label{eq:ch5:heis2_c2}
\cop(t) = \cop(T)\ee^{-\ii\omega (t-T)}-\ii g\int\limits_T^t \dd{t^\prime}\ee^{-\ii\omega(t-t^\prime)}\left(\aop^\dagger(t^\prime)+\aop(t^\prime)\right)
\end{equation} 
We can now integrate $\cop(t)$ over all frequencies. Using definition  \eqref{eq:ch5:heis2_c} or  \eqref{eq:ch5:heis2_c2} we get respectively:
\begin{subequations}
\begin{align}
\int c(t) \dd{\omega} &= \int  \dd{\omega} c(0)\ee^{-i\omega t} -\ii g\int\limits_0^t\dd{t^\prime}\left(\aop^\dagger(t^\prime)+\aop(t^\prime)\right)\int \dd{\omega}\ee^{-\ii\omega(t-t^\prime)}\notag \\
&= -\sqrt{2\pi} c_\mathrm{in}(t) - i\pi g \left(\aop^\dagger(t)+\aop(t)\right)\\
\int c(t) \dd{\omega} &= \hunderbrace{\int  \dd{\omega} c(T)\ee^{-i\omega (t-T)}}{\sqrt{2\pi} c_\mathrm{out}} -\ii g\int\limits_T^t\dd{t^\prime}\left(\aop^\dagger(t^\prime)+\aop(t^\prime)\right)\int \dd{\omega}\ee^{-\ii\omega(t-t^\prime)}\notag \\
&= \sqrt{2\pi} c_\mathrm{out}(t) + i\pi g \left(\aop^\dagger(t)+\aop(t)\right)
\end{align}
\end{subequations}
where we defined the input mode of the optical field just after the interaction as:
\begin{equation}\label{eq:ch5:cout}
\cop_\mathrm{out}(t) = \sqrt{\frac{1}{2\pi}}\int \dd{\omega} \cop(0)\ee^{-i\omega t}
\end{equation}
Imposing continuity at the interaction time, we finally get the output mode as a function of the input optical mode and the interaction with the mechanical system:
\begin{equation}
\cop_\mathrm{out}(t) = \cop_\mathrm{in}(t)-i\sqrt{\Gamma_\mathrm{ba}}\left(\aop^\dagger(t)+\aop(t)\right)
\end{equation}
In \textcolor{OliveGreen}{\st{real}} experiments however we likely will not be able to detect the output field with unity efficiency. Losses due to imperfect mode matching between the the optical output mode and the mode we detect, or due to non unity conversion efficiency of the photo detectors can me modelled by a single beam splitter with transmissivity $\eta_d$.
%\marginpar{ \includegraphics[width=2.5cm]{images/vacuum}}
Normalization of the detected mode function is ensured by considering a fraction $1-\eta_d$ of the optical vacuum coupling in through the orthogonal port. We therefore redefine the optical detected mode as:
%\marginpar{Properties:\\$\cop_\mathrm{vac}\cop_\mathrm{vac}=\cdag_\mathrm{vac}\cop_\mathrm{vac}=\cdag_\mathrm{vac}\cdag_\mathrm{vac}=0$\\$\cop_\mathrm{vac}\cdag_\mathrm{vac}=1$\\$\cop_\mathrm{vac}\cop_\mathrm{in}=\cdag_\mathrm{vac}\cop_\mathrm{in}=\cdag_\mathrm{vac}\cdag_\mathrm{in}=\cop_\mathrm{vac}\cdag_\mathrm{in}=0$}
\begin{equation}\label{eq:ch5:cdet}
\begin{split}
\cop_\mathrm{det}(t) &= \sqrt{1-\eta_\mathrm{d}}\cop_\mathrm{vac}(t) + \sqrt{\eta_\mathrm{d}}\cop_\mathrm{out}(t)\\
&=\sqrt{1-\eta_\mathrm{d}}\cop_\mathrm{vac}(t)+ \sqrt{\eta_\mathrm{d}}\cop_\mathrm{in}(t)-i\sqrt{\Gamma_\mathrm{meas}}\left(\aop^\dagger(t)+\aop(t)\right)
\end{split}
\end{equation}
where $\Gamma_\mathrm{meas} = \eta_\mathrm{d}\Gamma_\mathrm{ba}=\eta_\mathrm{d}2\pi g^2 $. 
Equation \eqref{eq:ch5:cout} (and \eqref{eq:ch5:cdet}) define the output (detected) optical field after the interaction with the mechanical oscillator. It depends on the input field and on the dynamics of the mechanical oscillator. As introduced in the main text, ponderomotive squeezing appears on timescales defined by the mechanical motion, much slower than the optical frequency. It is thus convenient to restrict the theoretical description to a frequency range close to the mechanical frequency. The natural way of operating such a frequency  cut-off is analyzing the fields in their spectral domain. Next, we will derive the motional spectrum of the position of the harmonic oscillator from equations \eqref{eq:ch5:aomega} in the high-Q and low frequency limit. We will then write the spectrum of the quadratures of the output (or detected) field.

\subsection{The motional spectrum of the quantum harmonic oscillator}

The position and momentum operators of the quantum mechanical oscillator can be written in terms of the $\aop$ and $\adag$ operators as
\begin{equation}
q = \hunderbrace{\sqrt{\frac{\hbar}{2m\Omega_q}}}{q_\mathrm{zpf}}\left(\aop^\dagger+\aop\right), \quad \mathrm{and} \quad p = \ii\hunderbrace{\sqrt{\frac{\hbar m \Omega_q}{2}}}{p_\mathrm{zpf}}\left(\aop^\dagger-\aop\right) 
\end{equation}
these definitions, together with equations \eqref{eq:ch5:aomega} allow us to write the motional spectrum of the harmonic oscillator. The computation is greatly simplified by observing that according to the correlators \eqref{eq:ch5:correlators} most terms vanish. The motional 
spectrum is~\cite{Hauer2015}:
%\marginpar{Definition:\\$S_{XX}(\omega) = \int \dd{\omega^\prime}\mean{X^\dagger(\omega)X(\omega^\prime)}$}
%\marginpar{\vspace{6cm}\\Approximations:\\$n_\mathrm{th} \gg 1$\\$\Omega_q\gg\gamma$\\$\omega \approx \Omega_q$\\$ $\\Definition:\\$\chi(\omega) = \frac{1}{m\left(\Omega_q^2-\omega^2 -i\gamma\omega\right)}$} 
\begin{equation}\label{eq:ch5:motionalpsd}
\begin{split}
&S_{qq}(\omega) = \int \dd{\omega^\prime} \mean{q(\omega)q(\omega^\prime)}\\ 
&\approx q_\mathrm{zpf}^2 \frac{\gamma n_\mathrm{th}4\Omega_q^2}{\left(\Omega_q^2-\omega^2\right)^2+\omega^2\gamma^2} + q_\mathrm{zpf}^2 \frac{\Gamma_\mathrm{ba} 4 \Omega_q^2}{\left(\Omega_q^2-\omega^2\right)^2+\omega^2\gamma^2}\\
%&=\left(q_\mathrm{zpf}^2 \gamma n_\mathrm{th} 4\Omega_q^2 + q_\mathrm{zpf}^2 \Gamma_\mathrm{ba} 4\Omega_q^2\right) m^2\lvert\chi_\mathrm{m}(\omega)\rvert^2 \\
%&= \left(2k_\mathrm{B}T\gamma m + 4 \Gamma_\mathrm{ba} p_\mathrm{zpf}\right) \lvert\chi_\mathrm{m}(\omega)\rvert^2\\ 
&= \left(S_{FF}^\mathrm{th}+S_{FF}^\mathrm{ba} \right)\lvert\chi_\mathrm{m}(\omega)\rvert^2,
\end{split}
\end{equation}
%\marginpar{\vspace{-11cm}\\ {Definition:\\$S_{XX}(\omega) = \int \dd{\omega^\prime}\mean{X^\dagger(\omega)X(\omega^\prime)}$}}
%\marginpar{\vspace{-4.5cm}\\Approximations:\\$n_\mathrm{th} \gg 1$\\$\Omega_q\gg\gamma$\\$\omega \approx \Omega_q$\\$ $\\Definition:\\$\chi_\mathrm{m}(\omega) = \frac{1}{m\left(\Omega_q^2-\omega^2 +i\gamma\omega\right)}$} 
with the meachnical susceptibility being $\chi(\omega) = [m\left(\Omega_q^2-\omega^2 -i\gamma\omega\right)]^{-1}$, $n_\mathrm{th} \gg 1$, $\Omega_q\gg\gamma$, $\omega \approx \Omega_q$,
Later in this chapter, when using these approximate results we will use $\Omega$, to indicate frequencies rather than $\omega$, to keep in mind that derivations are valid in a limited frequency range close to the mechanical resonance.

\subsection{The spectrum of the output quadratures}

Finally, we analyze the properties of the quadrature output fields, which one can detect experimentally by homodyne or heterodyne detection. In the following steps we will discuss for simplicity the properties of the output field, defined by  \eqref{eq:ch5:cout}. The derivation for spectrum of the quadratures of the detected mode \eqref{eq:ch5:cdet}, follows exactly the same steps, so that only the final result is given here.     
The general quadrature of angle $\theta$ of the optical field is defined as:
\begin{equation}
X^{\mathrm{j}}_{\theta} = \cop_\mathrm{j}\ee^{-\ii\theta} + \cop^\dag_\mathrm{j}\ee^{\ii\theta}
\end{equation}
Using equation \eqref{eq:ch5:cout} the quadrature of the output mode is:
\begin{equation}\label{eq:ch5:quad2}
\begin{split}
X^{\mathrm{out}}_{\theta}(t)&=\cop_\mathrm{out}\ee^{-\ii\theta} + \cop^{\dagger}_\mathrm{out}\ee^{\ii\theta}\\
&=\cop_\mathrm{in}\ee^{-\ii\theta} + \cop^{\dagger}_\mathrm{in}\ee^{\ii\theta} +\ii \sqrt{\Gamma_{\mathrm{ba}}}\left(\aop^\dagger(t)+\aop(t)\right)\left(\ee^{\ii \theta}-\ee^{-\ii\theta}\right)\\
&=X^{\mathrm{in}}_{\theta}(t)- 2\frac{\sqrt{\Gamma_{\mathrm{ba}}}}{q_{\mathrm{zpf}}}q(t)\sin(\theta)
\end{split}
\end{equation}
%\marginpar{\vspace{-2cm}\\Math reminder:\\$\frac{\ee^{\ii \theta}-\ee^{-\ii\theta}}{2\ii}=\sin(\theta)$}
and its autocorrelation function is:
\begin{equation}\label{eq:ch5:quadcorrelator}
\begin{split}
\mean{X^{\mathrm{out}}_{\theta}(t)^* X^{\mathrm{out}}_{\theta}(0)}&= \hoverbrace{\mean{X^{\mathrm{in}}_{\theta}(t)^\dagger  X^{\mathrm{in}}_{\theta}(0)}}{A2}+\hoverbrace{\frac{4\Gamma_\mathrm{ba}}{q_\mathrm{zpf}^2}\sin^2(\theta)\mean{q(t)^\dagger q(0)}}{B2} \\ 
&+\hunderbrace{\frac{2\sqrt{\Gamma_\mathrm{ba}}}{q_\mathrm{zpf}}\sin(\theta)\left[\mean{X^{\mathrm{in}}_{\theta}(t)^\dagger q(0)}+\mean{q(t)^\dagger X^{\mathrm{in}}_{\theta}(0)}\right]}{C2}
\end{split}
\end{equation}
Together with the Wiener-Khinchin theorem, the power spectral density of the output quadrature can be determined.
%\marginpar{Wiener-Khinchin theorem:\\$\int \dd{t} \mean{x(t)^\dagger x(0)}\ee^{-\ii\omega t}= S_{xx}(\omega) $} 
We compute the spectral contributions for the separate terms. The term $A2$ results in the trivial white spectrum:
\begin{equation}\label{eq:ch5:sa2}
S_{A2} = S_{XX}^{in}=\int\dd{t}\ee^{-i\omega t} \mean{X^{\mathrm{in}}_{\theta}(t)^\dagger  X^{\mathrm{in}}_{\theta}(0)} =\int\dd{\tau}\ee^{i\omega\tau} \delta(t) = 1
\end{equation}
the term $B2$, represents autocorrelation the position of the mechanical oscillator imprinted on the output field. Its spectrum is:
\begin{equation}\label{eq:ch5:sb2}
S_{B2} = S_{qq}^\mathrm{out}(\omega,\theta) = \frac{4\Gamma_\mathrm{ba}}{q_\mathrm{zpf}^2}\sin^2(\theta)\int\dd{t}\ee^{-i\omega t} \mean{q(t)^\dagger q(0)} =  \frac{4\Gamma_\mathrm{ba}}{q_\mathrm{zpf}^2}\sin^2(\theta) S_{qq}(\omega)
\end{equation}
where the motional spectral density is derived in equation \eqref{eq:ch5:motionalpsd}.
The last term ($C2$) in equation \eqref{eq:ch5:quadcorrelator} is the one showing the correlations between the amplitude fluctuations of the input field and the motion of the oscillator: the squeezing term. To derive its power spectrum it is convenient to write the input field as linear combination $\cos(\theta)$ and $\sin(\theta)$:
\begin{equation}\label{eq:ch5:quadratures}
X^{\mathrm{j}}_{\theta}  =\cop_\mathrm{j}\ee^{-\ii\theta} +\cdag_\mathrm{j}\ee^{\ii\theta} = X^{\mathrm{j}}_{0}\cos(\theta) + X^{\mathrm{j}}_{\frac{\pi}{2}}\sin(\theta)
\end{equation}
where the coefficients $X^{\mathrm{j}}_{0} = \cdag_\mathrm{j}+\cop_\mathrm{j}$ and $X^{\mathrm{j}}_{\frac{\pi}{2}} = i(\cdag_\mathrm{j}-\cop_\mathrm{j})$ represent respectively the amplitude and phase quadrature of the field. This decomposition leads to an immediate identification of the Hermitian and anti-Hermitian components of the quadrature. Because the $q$ operator is also Hermitian, one sees immediately that all terms proportional to $X^{\mathrm{j}}_{\frac{\pi}{2}}$ vanish in $C2$ equation \eqref{eq:ch5:quadcorrelator}, leaving only:
%\marginpar{Math reminder:\\$2\sin(\theta)\cos(\theta) = \sin(2\theta)$}
\begin{equation}\label{eq:ch5:c2}
C2 = \frac{2\sqrt{\Gamma_\mathrm{ba}}}{q_\mathrm{zpf}}\sin(2\theta)\mean{X^{\mathrm{in}}_{0}(t) q(0)} = \frac{2}{\hbar}\sin(2\theta)\mean{F^{\mathrm{ba}}(t),q(0)}
\end{equation}
Here we  have defined, for ease of physical interpretation, the radiation pressure (or ponderomotive) force determined by the optical amplitude fluctuations on to the mechanical oscillator as: \cite{SafaviNaeini2013}
\begin{equation}\
F^{\mathrm{ba}}(t) =  \frac{\hbar\sqrt{\Gamma_{\mathrm{ba}}}}{q_{\mathrm{zpf}}}
X^{\mathrm{in}}_{0}(t)
\end{equation}
From equation \eqref{eq:ch5:c2} the $C2$ term in \eqref{eq:ch5:quadcorrelator} represents the correlations between the input optical amplitude, defining a ponderomotive force via radiation pressure, and the position of the oscillator.
Recalling definitions \eqref{eq:ch5:aomega} and \eqref{eq:ch5:correlators}, the spectrum of $C2$ can be calculated  (using the convolution theorem)
%\marginpar{$\cop_\mathrm{in}\cop_\mathrm{in}=\cdag_\mathrm{in}\cop_\mathrm{in}=\cdag_\mathrm{in}\cdag_\mathrm{in}=0$\\$\cop_\mathrm{in}\cdag_\mathrm{in}=1$\\$ $\\$\adag\cdag_\mathrm{in} = -\frac{\sqrt{\Gamma_\mathrm{ba}}\cop_\mathrm{in}\cdag_\mathrm{in}}{-\ii\left(\Omega_q+\omega\right)+\frac{\gamma}{2}} $\\$ $\\$\adag\cop_\mathrm{in} = 0$\\$ $\\$\adag\cdag_\mathrm{in} = \frac{\sqrt{\Gamma_\mathrm{ba}}\cop_\mathrm{in}\cdag_\mathrm{in}}{\ii\left(\Omega_q-\omega\right)+\frac{\gamma}{2}} $\\$ $\\$\aop\cop_\mathrm{in}=0$}:
\begin{equation}\label{eq:ch5:sc2}
\begin{split}
S_{C2} = S_{qX}^\mathrm{out}(\omega,\theta) &= \frac{2\sqrt{\Gamma_\mathrm{ba}}}{q_\mathrm{zpf}}\sin(2\theta) \int \dd{t} \ee^{-i\omega t} \mean{X^{\mathrm{in}}_{0}(t) q(0)}\\
&=\frac{2\sqrt{\Gamma_\mathrm{ba}}}{q_\mathrm{zpf}} \sin(2\theta) X^{\mathrm{in}}_{0}(\omega) q(\omega)\\
&= \frac{2\sqrt{\Gamma_\mathrm{ba}}}{q_\mathrm{zpf}} \sin(2\theta) (\cdag_\mathrm{in} +\cop_\mathrm{in})q_\mathrm{zpf}(\adag+\aop)\\
%&= 2\Gamma_\mathrm{ba} \sin(2\theta)\ii\left(\frac{\cop_\mathrm{in}\cdag_\mathrm{in}}{\ii \left( \Omega_q - \omega \right) +\frac{\gamma}{2}} - \frac{\cop_\mathrm{in}\cdag_\mathrm{in}}{-\ii \left( \Omega_q\approx + \omega \right) +\frac{\gamma}{2}}\right)\\
%&\approx 4\Gamma_\mathrm{ba} \sin(2\theta) \Omega_q m \chi_\mathrm{m}^*(\omega)\\
&\approx\frac{4\Gamma_\mathrm{ba}}{q^2_\mathrm{zpf}} \frac{\hbar}{2}\chi_\mathrm{m}^*(\omega)\sin(2\theta)
\end{split}
\end{equation}
where again we have used the high-Q approximation, in the vicinity of resonance.
Putting together \eqref{eq:ch5:sa2}, \eqref{eq:ch5:sb2} and \eqref{eq:ch5:sc2}, the real part of the total output spectrum of the output quadrature becomes:
\begin{equation}
S_{X,X}^{\mathrm{det}}(\Omega,\theta) = S_{\theta,\theta}^{\mathrm{det}}(\Omega) = 1 + \frac{4\Gamma_{\mathrm{ba}}}{z_{\mathrm{zpf}}^2} 
\left[S_{qq}(\Omega)\sin^2(\theta)+\frac{\hbar}{2}\mathrm{Re}\left\lbrace\chi_\mathrm{m}(\Omega)\right\rbrace\sin(2\theta)\right]
\end{equation}
This is, as expected, the same result that one gets if considering a mechanical oscillator coupled to an optical resonator, driven on resonance and in the bad cavity limit\cite{SafaviNaeini2013}.
Note that for simplicity we have operated a change of notation in the representation of the cross spectrum: we indicate the quadratures involved by their defining angle $\theta$ in place of the full quadrature name $X_{\theta}$.
As mentioned previously, the spectrum of the detected quadratures in the case of non unity detection efficiency, can be derived by following the same steps, starting from the definition of $\cop_\mathrm{det}$ in \eqref{eq:ch5:cdet} rather than using $\cop_\mathrm{out}$ from \eqref{eq:ch5:quad2}, and using the fact that the vacuum and the input field are uncorrelated. The result is:
%\marginpar{ \includegraphics[width=3.5cm]{images/chi2}}
%\marginpar{ \includegraphics[width=3.5cm]{images/rechi}}
%\marginpar{ \includegraphics[width=3.5cm]{images/sxxponder}}
\begin{equation}\label{eq:ch5:ponderomotive1}
\begin{split}
S_{\theta,\theta}^{\mathrm{det}}(\Omega) &= 1 + \frac{4\Gamma_{\mathrm{meas}}}{q_{\mathrm{zpf}}^2} 
\left[S_{qq}(\Omega)\sin^2(\theta)+\frac{\hbar}{2}\mathrm{Re}\left\lbrace\chi_\mathrm{m}(\Omega)\right\rbrace\sin(2\theta)\right]\\
& = 1 + \frac{4\eta_\mathrm{d}\Gamma_{\mathrm{ba}}}{q_{\mathrm{zpf}}^2} 
\left[\left(1+\frac{1}{C_\mathrm{q}}\right) S_{FF}^\mathrm{ba}\lvert\chi_\mathrm{m}(\Omega)\rvert^2 \sin^2(\theta)+\frac{\hbar}{2}\mathrm{Re}\left\lbrace\chi_\mathrm{m}(\Omega)\right\rbrace\sin(2\theta)\right]
\end{split}
\end{equation}
In this definition, the measured spectrum is given in units of the detected shot noise, first term of the equation. The second term in \eqref{eq:ch5:ponderomotive1} takes into account the motion of the harmonic oscillator transduced to the output quadrature, and is always positive. The third term, on the other hand, is the result of correlations between the motion of the particle and the input filed, which for certain frequency and quadrature angles may be negative. This may lead, in certain conditions, to a total noise that is smaller than the optical vacuum fluctuations. The optical mode at that frequency is therefore a squeezed optical state. 

Let us have a closer look at the behavior of this function with frequency and quadrature angle. Equation \eqref{eq:ch5:ponderomotive1} is a function of two variables, defined on the domain $\Omega \in [0,\infty)$ and $\theta \in [-\pi/2,\pi/2]$. In this domain the function is maximized in $(\Omega_q,\pm \pi/2)$, and has two local minima at $(\Omega_q + \Delta_\mathrm{sq},\theta_\mathrm{sq})$ and $(\Omega_q - \Delta_\mathrm{sq},-\theta_\mathrm{sq})$. At $(\Omega_q ,0)$ is a saddle point. The two minima represent two frequency modes $\Omega_\mathrm{sq} = \Omega_q \pm \Delta_\mathrm{sq}$ where light is maximally squeezed, respectively along a specific angle $\pm\theta_\mathrm{sq}$. 
\begin{figure}[h]
\centering
\includegraphics[scale=1]{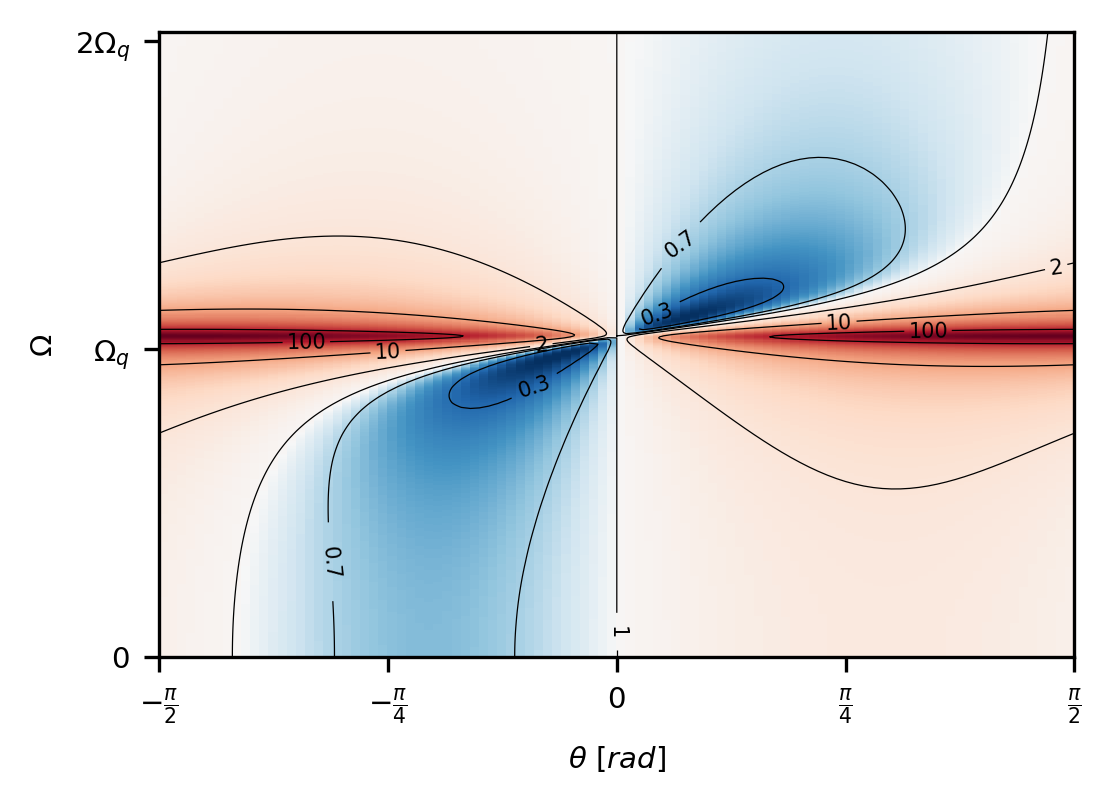}
\caption{\textbf{Theoretical spectrogram.} The spectrogram of the output quadrature as a function of the angle of analysis, evaluated for $C_\mathrm{q}\gg1$, and $\eta_\mathrm{d}=1$.}
\label{fig:ponderomotivetheory}
\end{figure}
In particular in the high-Q ($\gamma\ll\Omega$) and high cooperativity ($C_\mathrm{q}\gg1$), with perfect detection efficiency $(\eta_\mathrm{d})=1$, and high occupancy $n\gg1$, the correlation term is strong enough, for two specific modes, to maximally suppress both the thermal noise and the shot noise contributions.
%$S_{XX}^\mathrm{ideal} (\Omega_q\pm\Delta_\mathrm{sq}, \pm \theta_\mathrm{sq}) \rightarrow 0$. And the squeezing parameter (defined as the inverse total noise in units of the vacuum noise) $s(\Omega) = 1/S_{XX}(\Omega, \pm \theta_\mathrm{sq}) \rightarrow \infty$. Considering non unity detection efficiency and a finite enviromental coupling, the  the squeezing parameter is reduced to
%\begin{equation}
%s  = \frac{1}{1-\eta_\mathrm{d}\left(1-\frac{1}{C_\mathrm{q}}\right)}
%\end{equation}
%where again $\eta$ represents the total information efficiency, and take into account both detection imperfection and environmental disturbances.
Equation \eqref{eq:ch5:ponderomotive1} is sometimes written in the units of the displacement spectrum.
%Remembering the definition of measurement rate given in chapter xx, equation \eqref{eq:ch1:rates} 
By multiplying equation \eqref{eq:ch5:ponderomotive1} by $S_{qq}^\mathrm{imp}$, we obtain the spectrum of the output quadrature in units of displacement noise~\cite{Mason2019}:
 \begin{equation}\label{eq:ch5:ponderomotive2}
S_{\theta,\theta}^{\mathrm{out}}(\Omega) = S^{\mathrm{imp}}_{qq} + \sin^2(\theta)
\left[ \left( S^{\mathrm{ba}}_{FF} + S^{\mathrm{th}}_{FF} \right) \vert\chi_\mathrm{m}(\Omega)\vert^2  
 -2 \mathrm{Re}\left\lbrace\chi_\mathrm{m}(\Omega)S_{qF}^{\mathrm{ba}}(\theta) \right\rbrace \right]
\end{equation}
where the force-displacement correlations spectrum is $S_{qF}^{\mathrm{ba}}(\theta) = -\frac{\hbar}{2}\cot(\theta)$.

\section{Experiment}

\subsection{Detection}

We reconstruct the orthogonal quadratures of the optical output field from a heterodyne measurement. The advantage of this method compared to a single homodyne detection is that it allows a full tomography of the state at all times and does not require any  interferometric stabilization. As a drawback, recording the heterodyne beat note, elastic component of the scattering process, requires a large detection dynamic range to allow amplification of the quantum correlations with a strong local oscillator without saturating the detector. The required dynamic range can be gauged by the inverse square of the Lamb-Dicke parameter $\eta_\mathrm{LD} =\alpha k z_\mathrm{zpf}$ ($\alpha$: geometric factor of order 1, $k$: wavevector, $z_\mathrm{zpf}=\sqrt{\hbar/(2m\Omega_z)}$: position zero-point fluctuation), defining the ratio of inelastic to elastic scattering, together with the total detection efficiency $\eta_\mathrm{d}$\footnote{ here we consider the total detection efficiency of the heterodyne measurement}. In order to reduce the required dynamic range we choose for this experiment a small particle size of $r=43\,\mathrm{nm}$, and low mechanical frequency of $\Omega_z/2\pi\sim 75\,\mathrm{kHz}$. In our case, with $\alpha\sim1.5$ and $\eta_\mathrm{d}\sim0.3$ we have:
\begin{equation}
  \eta_\mathrm{d}/\eta_\mathrm{LD}^{2}\sim 2.1\cdot10^7
\end{equation}
Which is still small enough when using a 16 bit oscilloscope which has a dynamic range of ${2^{16}}^2=4.2\cdot10^9$. 
In addition it is important to choose the appropriate detector: in order to achieve a shot noise limited detection one must be able to amplify the signal light with a local oscillator that is strong enough to overcome the detectors dark noise, without damaging the photodiodes nor saturating the transimpedance amplifier:
\begin{equation}\label{lo}
P_\mathrm{lo} < \min\left\{
2P_\mathrm{max}, \frac{\Delta P_\mathrm{sat}^2}{P_\mathrm{sig}}\right\}
\end{equation}
where $P_\mathrm{max}$ is the single photodiode damage power, $\Delta P_\mathrm{sat}$ is the power difference between the photodiodes saturating the detector, and $P_\mathrm{sig}$ the power of the signal that is being heterodyned. Sometimes the saturation power is  in terms of the maximum voltage swing ($\Delta V$) at the output of the detector:
\begin{equation}\label{dv}
\Delta P_\mathrm{sat} =   \frac{-e\eta_\mathrm{q}}{h\nu}\frac{\Delta V}{g_\mathrm{t}}  
\end{equation}
where $g_\mathrm{t}$ is the detectors transimpedance gain, $e$ is the electron charge, $\eta_\mathrm{q}$ the detector quantum efficiency, $h$ the Planck constant and $\nu$ the optical frequency.
In terms of the induced photocurrent we will have a shot noise and a dark noise level of respectively (in units of $\mathrm{A}/\sqrt{\mathrm{Hz}}$):
\begin{equation}\label{noise}
\sqrt{S_i^\mathrm{sn}}= -e\sqrt{2\eta_\mathrm{q}\frac{P_\mathrm{lo}}{h\nu}} \quad \mathrm{and} \quad \sqrt{S_i^\mathrm{dn}} = \mathrm{NEP}\frac{\eta_\mathrm{q}e}{h\nu},
\end{equation}
with NEP the noise equivalent power of the detector. The ratio of optical shot noise to dark noise is the figure of merit that has to be maximized in the choice of a detector. Putting together definitions \eqref{lo}, \eqref{dv} and \eqref{noise} we find:
\begin{equation}
   \frac{S_i^\mathrm{sn}}{S_i^\mathrm{dn}} = \frac{2h\nu P_\mathrm{lo}}{\eta_q \mathrm{NEP}^2} < \min\left\{\frac{4h\nu}{\eta_q}
\frac{ P_\mathrm{max}}{ \mathrm{NEP}^2}, \frac{2h^3\nu^3}{e^2\eta_\mathrm{q}^2} \frac{ \Delta V^2}{ g_\mathrm{t}^2\mathrm{NEP}^2P_\mathrm{sig}}\right\}
\end{equation}

\begin{figure}[h!]
\includegraphics[scale=1]{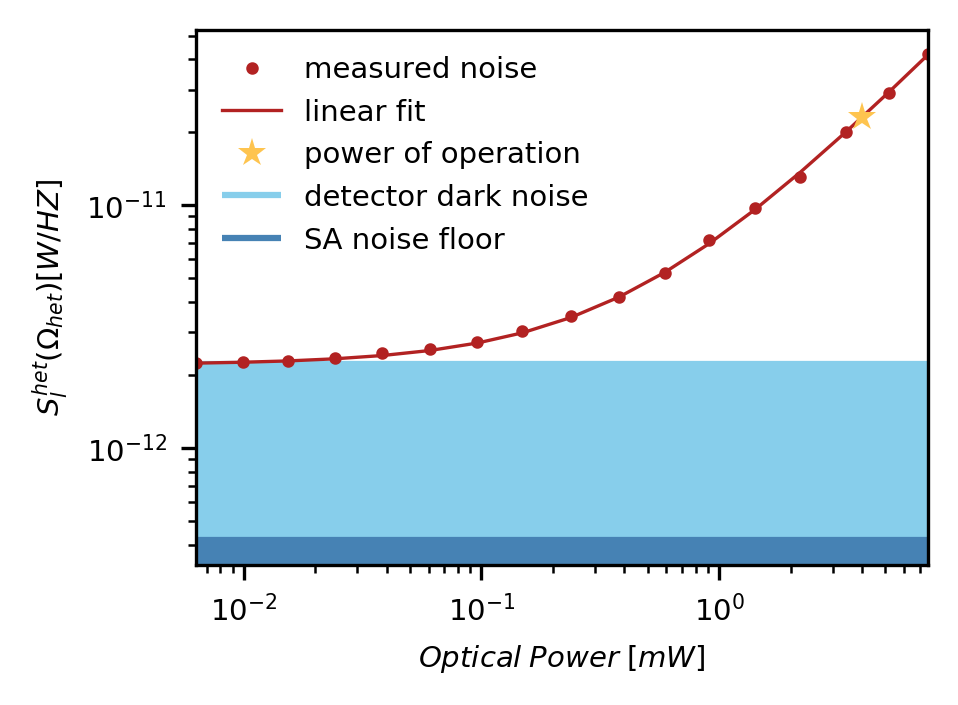}
\caption{\textbf{Noise contributions}: We operate our measurement at a local oscillator power of $4\,\mathrm{mW}$. In these conditions we achieve a shot noise level 10 dB above dark noise without saturating the detector.}
\label{fig:shotnoise}
\end{figure}

We use a Thorlabs PDB 471 balanced detector, with a maximum NEP of $8\, \mathrm{pW}/\sqrt{\mathrm{Hz}}$ a maximum voltage swing of $\Delta V \pm 3.6 \, \mathrm{V}$, a transimpedance gain of $g_\mathrm{t} = 10^4$ and a maximum power per diode of $P_\mathrm{max} = 5\,\mathrm{mW}$. With a signal power of only $P_\mathrm{sig} \sim 100\,\mathrm{nW}$ we are limited by the power threshold of the diodes to maximum shot noise level of about 15 dB above dark noise. In practice operating below threshold and considering digitization noise we measure our shot noise 11 dB above dark noise (see Figure \ref{fig:shotnoise}). 

\subsection{Heterodyne demodulation}

The output field (oscillating at frequency $\omega_\mathrm{in}$) is interfered on a 50/50 beam splitter with a reference field (local oscillator) oscillating at a frequency $\omega_\mathrm{lo} = \omega_\mathrm{in}+\Omega_\mathrm{het}$. The intensity at two outputs is then recorded by two photodiodes, resulting in a photocurrent difference of:
\begin{equation}
i_\mathrm{het} \propto  X_0^{out}(t)\cos(\Omega_\mathrm{het} t + X_{\frac{\pi}{2}}^{out}(t) + \varphi(t))
\end{equation}
where $\varphi(t)$ is a slowly  varying phase ($\dd\varphi(t)/\dd t \ll \Omega_z, \Omega_\mathrm{het}$ $\forall t$ ) that takes into account the drifts in the optical path length.
If the phase scan (or heterodyne frequency) is much faster than the frequencies of the dynamics of interest, a demodulation of the heterodyne signal by $\sin(\Omega_\mathrm{het} t)$ and $\cos(\Omega_\mathrm{het} t)$ will result in a simultaneous recording of the in-phase ($I$) and quadrature ($Q$) components of the signal. From these, a coordinate change and equation \eqref{eq:ch5:quadratures}, allow to reconstruct the time evolution of any quadrature $X_\theta^{out}(t)$.
The demodulation can easily done either with analog electronic components, by use of an IQ mixer and low pass filters, or even digitally using a Hilbert transform signal decomposition. 
As the phase drifts in the local oscillator are completely recorded in the heterodyne beat note, the reconstruction of the orthogonal quadratures does not require active interferometric stabilization.

\subsection{Evaluation of the system parameters}

%\begin{equation}
%S_{\theta,\theta}(\omega) =  1 + \frac{4\Gamma_{\mathrm{meas}}}{q_{\mathrm{zpf}}^2} 
%\left[4p_\mathrm{zpf}^2\Gamma^\mathrm{tot}\lvert\chi_\mathrm{m}(\Omega)\rvert^2 %\sin^2(\theta)+\frac{\hbar}{2}\mathrm{Re}\left\lbrace\chi_\mathrm{m}(\Omega)\right\rbrace\si%n(2\theta)\right]. 
%\end{equation}

The evaluation of the measurement rate $\Gamma_{\mathrm{meas}}$ and of the total decoherence rate $\Gamma_{\mathrm{tot}}$ by fitting spectra of the optical quadratures is not a trivial task as the two parameters are strongly correlated.
In order to reduce the problem to a single parameter estimation, we take advantage once again of our heterodyne measurement, which also allows us to compute the cross-spectrum of two simultaneous quadratures. This gives us access to an absolute measurement of the steady state energy of the oscillator in units of its quanta of excitation. Knowing the total cooling rate (damping) and the steady state energy (occupation) we can finally estimate the total decoherence rate using the relation:
\begin{equation}\label{eq:heating}
n = \frac{\Gamma_\mathrm{tot}}{\gamma}   
\end{equation}

\subsubsection{Occupation}

To determine the total occupation of our system we consider the cross correlation spectrum of the output field quadratures  $X_\frac{\pi}{4}$ and $X_\frac{3\pi}{4}$:
\begin{equation}
S_{\frac{\pi}{4},\frac{3\pi}{4}}(\omega) = 2D\left(n+\frac{1}{2}\right)\operatorname{Im}\{\chi_m(\omega)\}+\ii D \operatorname{Im}\{\chi_m(\omega)\}
\end{equation}
with the constant $D = 4\Gamma_\mathrm{meas}\Omega_q m$. The ratio of the real and imaginary part of this cross correlation gives us a calibration-free estimate of the mean energy of the harmonic oscillator in units of $\hbar\Omega_q$~\cite{Purdy2017}:
\begin{equation}
    \frac{\operatorname{Re}\{S_{\frac{\pi}{4},\frac{3\pi}{4}}(\omega)\}}{\operatorname{Im}\{S_{\frac{\pi}{4},\frac{3\pi}{4}}(\omega)\}} = 2n+1
\end{equation}
From the ratio of the fitted peak heights~\cite{Purdy2017}, constraining all other parameters to be the same, we measure a total occupation of $n = 84.9 \pm 0.7$
\begin{figure}[h!]
\includegraphics[scale=1]{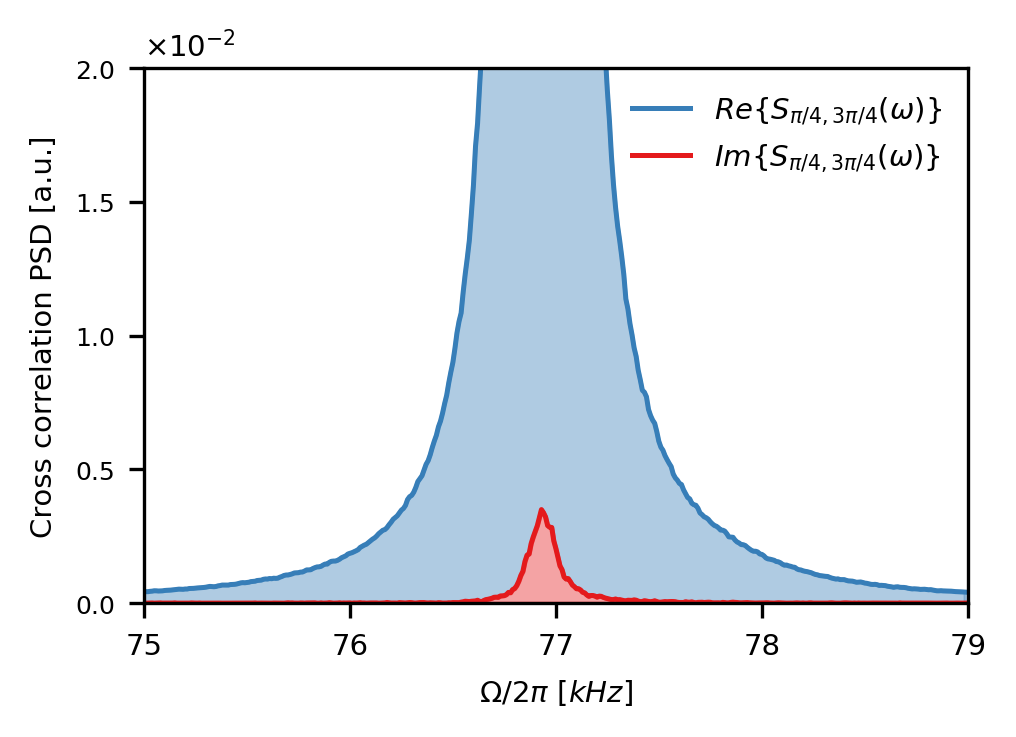}
\caption{\textbf{Estimation of the occupation.} Real (blue) and imaginary (red) spectra of the cross correlation $S_{\frac{\pi}{4},\frac{3\pi}{4}}$.The amplitude of the imaginary part is determined by the zero point fluctuation while the real part is proportional to the thermal occupation of the oscillator.}
\label{fig:thermometry1}
\end{figure}

\subsubsection{Damping}

\begin{figure}[h!]
\includegraphics[scale=1]{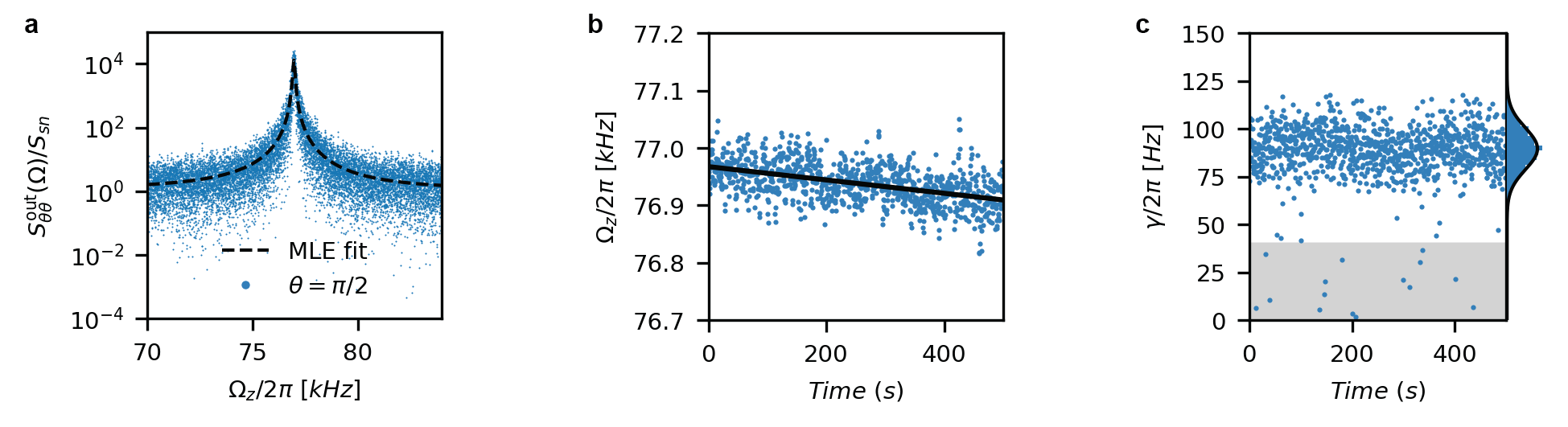}
\caption{\textbf{Estimation of the total damping and resonance frequency.} (a) PSD of the autocorrelation $S_{\frac{\pi}{2},\frac{\pi}{2}}$ averaged for 0.5 seconds and fitting for the estimation of $\Omega_{z}$. (b) Time analysis of the resonance frequency over the whole time of the measurement (500 s). A slow drift on the resonance frequency is observed during the measurement. (c) Fitted $\gamma$ every 0.5 second. While some fitted values are clearly out of the mean trend, a Gaussian distribution is obtained, enabling to estimate a mean value for the damping. The drift in the resonance frequency and low averaging over time hinders the fitting algorithm to estimate the linewidth.}
\label{fig:damping}
\end{figure}

The damping of the system defines the linewidth of the motional resonance peak. However, during the course of our measurements (500 s) drifts in the resonance frequency also result in a broadening of the averaged spectral line. This is due to slow thermomechanical drifts which in turn result in phase drifts of stray reflections into our trap, which (weakly) interfere with the trapping field, thus shifting the mechanical resonance frequency. In order to exclude this broadening from our damping estimation we split our trace into 0.5 second segments and fit to the linewidth and resonance frequency of the autocorrelation spectrum $S_{\frac{\pi}{2},\frac{\pi}{2}}$ where the squeezing term vanishes. The reason for fitting the auto-correlation spectrum is the high peak signal to noise ratio and background level dominated by shot noise. We fit the model using maximum likelihood estimation which takes into account the $\chi^2$ distribution of the non-averaged spectra. From these fits we can plot the time series of the fitted maximum likelihood parameters and observe indeed a slow drift ($\sim 7 \,\mathrm{Hz/min}$) in the mechanical frequency, while the linewidth remains stable over time (Figure \ref{fig:damping}). The measured damping is $\gamma/2\pi = 89.6\pm 0.3\,\mathrm{Hz}$.

\subsubsection{Total Heating rate}

The measured cooling rate, together with the steady state energy of the particle allow us to extrapolate the total heating, or decoherence rate. Inverting equation \eqref{eq:heating} we obtain a total heating rate of $\Gamma_\mathrm{tot}/2\pi = 7.6\pm0.1\,\mathrm{kHz}$

\subsubsection{Measurement rate}

We can now write equation \eqref{eq:ch5:ponderomotive1} for $\theta = \pi/2$:
\begin{equation}\label{psd_pi2pi2}
    S_{\frac{\pi}{2},\frac{\pi}{2}}(\omega) = 1+ 16m^2\Omega_q^2\Gamma_\mathrm{meas}\Gamma_\mathrm{tot}\lvert\chi_\mathrm{m}(\Omega)\rvert^2
\end{equation}
where we have used the relation $\Gamma_\mathrm{tot}= 4p_\mathrm{zpf}^2 S_{FF}^\mathrm{tot}$. Having fixed $\Gamma_\mathrm{tot}$ we can now fit equation \eqref{psd_pi2pi2} to the corresponding measured spectrum with $\Gamma_\mathrm{meas}$ as the only one free parameter.
We fit a total measurement rate of $\Gamma_\mathrm{meas}/2\pi  =0.6 \pm 0.1\, \mathrm{kHz}$.At a pressure of $1.5\cdot10^{-8}\,\mathrm{mbar}$ we estimate a thermal decoherence rate of $\Gamma_\mathrm{th}/2\pi = 2.7, \mathrm{kHz}$, resulting in a measurement cooperativity of $C_\mathrm{q}=\Gamma_\mathrm{ba}/\Gamma_\mathrm{th} = 1.8$, and a detection efficiency per quadrature of $\eta_\mathrm{d} = 0.11$.
\begin{figure}[h!]
\includegraphics[scale=1]{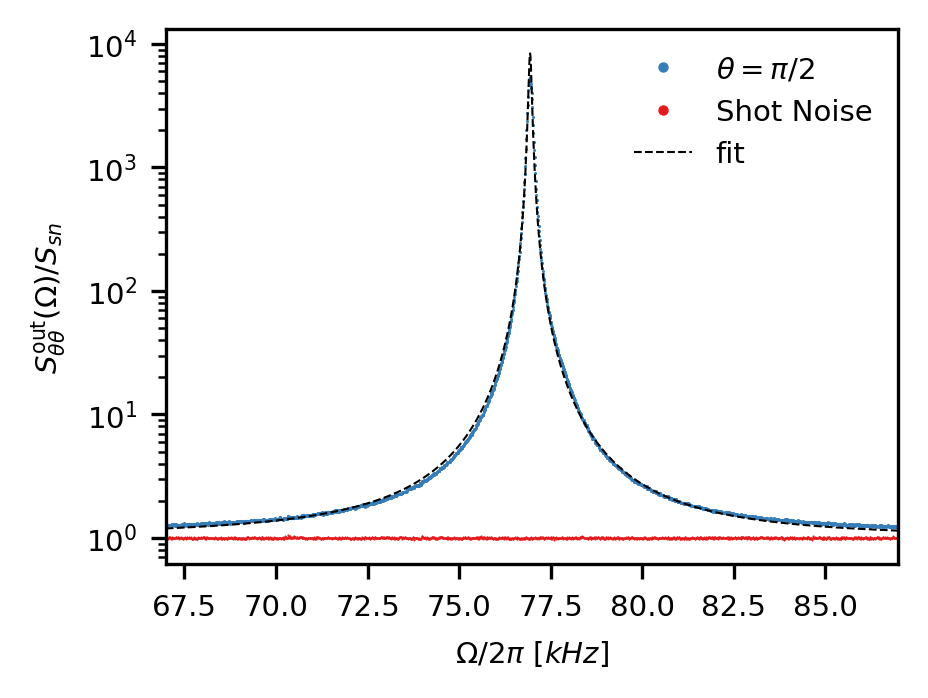}
\caption{\textbf{Estimation of the measurement rate.} PSD of the autocorrelation $S_{\frac{\pi}{2},\frac{\pi}{2}}$ averaged for the whole measured time (500 s). The quadrature angle $\pi/2$ is chosen to estimate $\Gamma_\mathrm{meas}$ as for this angle equation \eqref{psd_pi2pi2} is simplified.}
\label{fig:thermometry2}
\end{figure}

\end{document}